A CLASS OF EXAMPLES DEMONSTRATING THAT "$P \neq NP$" IN THE "P VS NP" PROBLEM

Vasil Penchev, vasildinev@gmail.com
Bulgarian Academy of Sciences: Dept. of Logical Systems and Models

*Abstract*: The CMI Millennium "P vs NP Problem" can be resolved e.g. if one shows at least one counterexample to the "$P = NP$" conjecture. A certain class of problems being such counterexamples will be formulated. This implies the rejection of the hypothesis "$P = NP$" for any conditions satisfying the formulation of the problem. Thus, the solution "$P \neq NP$" of the problem in general is proved. The class of counterexamples can be interpreted as any quantum superposition of any finite set of quantum states. The Kochen-Specker theorem is involved. Any fundamentally random choice among a finite set of alternatives belong to "NP' but not to "P". The conjecture that the set complement of "P" to "NP" can be described by that kind of choice exhaustively is formulated.

*I Introduction*

Let the problem be: "is Schrödinger's cat alive or dead?" One should guess its "state" in advance, before any actual observation.

No Turing machine can resolve it in advance in principle, for if it could do this, hidden variables would exist necessarily in quantum mechanics. However, the Kochen-Specker theorem (1967) states that the latter is not the case.

Well, now: the door of the cat's cell is open, and the Turing machine can check very quickly whether the cat is alive or not. Thus, "NP" (the latter checking time, "very quickly") is different from "P" (the former infinite time necessary for any Turing machine to guess in advance whether the cat is alive or not). The utilized metaphor of „Schrödinger's cat" is directed triply: it means immediately a quantum superposition, but furthermore a "fundamentally random choice" (defined rigorously bellow) as well as the work of a quantum computer in coherent state.

Informally, a fundamentally random choice has to be distinguished from "random choice" in common sense such as "coin flipping" or "playing dice". Both examples are deterministic in fact. There exist "hidden variables" remaining unknown as usual, but absolutely and exactly quantitatively knowable in principle. By inputting them in a computer, it will be able to calculate unambiguously the result of such an ostensibly random choice.

Thus, "flipping coin" and "Schrödinger's cat" are fundamentally different as to a choice claiming to be random. A nonempty finite set of hidden variables is available in the former, but not in the latter, and this distinguishes them. The Kochen – Specker theorem (1967) states the absence of hidden variables in quantum mechanics at all and allows for a rigorous definition of "fundamentally random choice" (which cannot be predetermined by any way in principle) in terms of "Turing machine": thus, it can be referred to the "P vs NP" problem directly.

Speaking figuratively, the "coin flipping" unlike "Schrödinger's cat" is not "random enough" to be able to offer a class of examples being "non-P, but NP".

Furthermore, the concept of quantum superposition (Dirac 1958: 14-18) represents the work of quantum computer in coherent state. Thus, if the class of "non-P, but NP" problems is defined by means of "quantum superposition", there exist calculations of quantum computer which cannot be simulate by any Turing machine in any polynomial time, but the solution of quantum computer can be checked by a Turing machine for a polynomial time[1].

---

[1] That statement does not mean the following two cases: (1) the simulation of quantum computer by a Turing machine for a non-polynomial, but finite time; (2) checking any result of quantum computer by a Turing for a polynomial time. Particularly, (1) there can exist calculations of quantum computer which cannot be simulate by any Turing machine for any finite time; (2) there can exist calculations of quantum computer which cannot be checked by any Turing machine for any finite time.

The calculation of quantum computer in coherent state and the definition of "absolutely random choice" are represented both by the same concept of "quantum superposition", and thus, by the same mathematical formalism based on the separable complex Hilbert space and its elements.

That equivalence (or isomorphism) implies the following problem needing rather a philosophical reflection further: is any "absolutely random choice" identical to a calculation of quantum computer in general or the relevant calculations of quantum computer are different class only intersecting with the class of absolutely random choices? That problem will not be discussed.

The exhibition bellow involves rather elementary mathematical tools and concepts such as: the axiom of choice and the well-ordering principle of set theory (e.g. ZFC) as well as their equivalence; "Turing machine"; the separable complex Hilbert space; the Kochen-Specker theorem (1967); the principle of superposition in quantum mechanics formulated mathematically by Dirac in 1930.

Peano arithmetic is not sufficient: the concept of actual infinity is utilized explicitly. However, no idea stronger than actual infinity is necessary.

Thus, the exhibition is thoroughly in the framework of the standard mathematics featured by Peano arithmetic and ZFC set theory.

The paper is structured as follows:

*Section II* justifies the specific concepts and notations further. *Section III* demonstrates that the class of problems formulated rigorously in Section II are "non-P problems". *Section IV* shows that the same class of problems are "NP" simultaneously and represents the results in the notations of *Official Problem Description*. *Section V* suggests two conjectures without proofs to outline a possible direction for future work.

Two "Appendixes" are added. *Appendix 1* reformulates the "Turing machine" as it defined in *Official Problem Description* in terms of the separable "$Z_2$" Hilbert space. The former does not involves "actual infinity" unlike the latter thus preparing the complete rigorous definition of quantum computer as a generalization of Turing machine in *Appendix 2*.

## II A general formulation of a class of problems demonstrating that "$P \neq NP$"

The following concepts specific for this paper are used further:

"Choice $(Ch)$" means a certain choice of a certain element of a set if nothing different is written or complemented expressively.

"Fundamentally random choice (FRC)": A choice of a certain element of a set is FRC if it can be accomplished only in virtue of the axiom of choice.

Furthermore, a choice of a certain element of a set can be accomplished by a finite number of criteria determining unambiguously a certain element of a set. That choice will be called "determined choice (DC)"

If a finite number of criteria determining unambiguously a certain element of a set exists necessarily, but those criteria are granted as unknown, the according choice is called "hidden determined choice (HDC)". The corresponding criteria (finite number) are called "hidden variables ($HV_{FC}$)" or "hidden parameters" meaning the certain values of $HV_{FC}$ for a certain element of a set to be chosen unambiguously.

The utilized term "hidden variables" coincides with the used in the paper of Kochen and Specker (1968). This equivalence is essential for the proof of the theorem above, and will be demonstrated by the proof of the *Statement 2*.

If only the existence of a finite number of criteria determining unambiguously a certain element of a set is proved, but those criteria cannot be explicated, the choice is called "random choice (RC)" respectively.

Furthermore, all definitions and notations referring to the "P vs NP problem" are the same as in the "Official Problem Description" supplied by CMI[2]: This means Stephen Cook's text entitled "P vs NP problem". All other definitions and notations necessary for the solution of the problem are introduced in the present text expressively. The proof meant further is thoroughly within the standard first order theory of mathematics, namely the ZFC set theory. This means that the intended proof is consistent to ZFC.

---

[2] https://www.claymath.org/sites/default/files/pvsnp.pdf

In fact, only a few, rather elementary tools belonging to the standard mathematics meant as above will be the separable complex Hilbert used explicitly further. One can enumerate them as follows: the principle of superposition in quantum mechanics (Dirac 1958: 14-18)[3]; the separable "$Z_2$"[5] Hilbert space, the Kochen – Specker theorem (1967) about the absence of hidden variables in quantum mechanics implied by the separable complex Hilbert space. Nothing else than all concepts involved expressively in the cited "Official Problem Description" is necessary.

A quantum system Q is described to be in a coherent state QS consisting of the superposition of a finite set S consisting of at least two elements of experimentally measurable results $R_n, n \geq 2$, e.g. an experiment about the electron spin or the photon polarization, etc. The coherent state is: $QS = (C_1, C_2, ..., C_n) = \sum_{k=1}^{n} C_k \, e^{ik\omega}, C_k \neq 0$ are complex numbers, $e^{ik\omega}$ is the k-th "axis" of the separable complex Hilbert space unambiguously corresponding to $R_k$.

The problem, further notated as "QP" belonging to the class of the problems demonstrating "$P \neq NP$" (further notated briefly as "$P \neq NP$ problems") is formulated so:

"Which result $R$ among $R_n$ will be measured after a certain measurement of Q?"

Let the measurement reduce QC to a certain result $R$ among $R_n$. That result R is determined unambiguously by a certain natural number $r \leq n$. The problem is:

Which is that "r"?

That problem does not belong to "P" for it cannot be resolved by a Turing machine in any finite time, and thus, for any polynomial time. The Kochen - Specker theorem (1967) excludes any hidden variable to exist for describing Q. An algorithm for Turing machine allowing for it to forecast the result of the measurement for any finite time is equivalent to the availability of hidden variables impossible according to the cited theorem. Thus, there is not any algorithm of that kind and particularly there does not exist any algorithm for Turing machine, able to resolve the problem for any polynomial time, "P".

Consequently, the problem QP belongs to the class of non-P problems (a rigorous proof in detail follows).

Simultaneously, it belongs to the class of NP problems: the TM has to resolve whether R is a certain element among the finite number n of elements of S, namely $R_n$.

Consequently: $\{QP \subset \neg P \land NP\} \rightarrow (P \neq NP)$.

*III The proof in detail that QP belongs to the class of non-P problems*
*Theorem*: Any problem defined as QP is not a P-problem (i.e. it is a non-P problem).
This statement will be proved as a corollary from the following:
*Theorem*: No Turing machine can resolve QP for any finite time.
*Corollary*: QP belongs to the class of non-P problems.
The specific premises of the proof are: the Kochen-Specker theorem, the equivalence of the axiom of choice (roughly speaking, an element can be chosen from any set) and the well-ordering theorem (or the "principle": any set can be ordered well). A terminological justification: "well-ordering postulate" will be used bellow in the same meaning as the well-ordering theorem as it is equivalent to the axiom

---

of a choice rather than a conclusion from it. "Well-ordering theorem" or "well-ordering principle" are rather misleading[4].

The pathway of the proof can be marked by a few intermediate statements:

*Statement 1*. QP implies no hidden variables in the rigorous meaning of "hidden variables" defined in the paper of Kochen and Specker (1967), or symbolically: $QP \rightarrow HV_{K\&S}$.

*Statement 2*. The meaning of the Kochen and Specker hidden variables is equivalent to a "finite set of unambiguous criteria (FC) for an element known as belonging to a certain set to be chosen from the set" (i.e. a different meaning of "hidden variables"); briefly: $HV_{K\&S} \leftrightarrow HV_{FC}$.

*Statement 3*. The finite set of unambiguous criteria for an element to be chosen from its set is equivalent to a determined choice and to the negation of a fundamentally random choice: $HV_{FC} \leftrightarrow DC \leftrightarrow \neg(FRC)$.

*Statement 3*.1. Any Turing machine (TM) can choose a certain element of a set only by DC (i.e. by a finite tuple of instructions), that is, $Ch_{TM} \leftrightarrow DC$.

*Statement 3.2*. Any parameter (i.e. a certain value of a certain HV) needs at least one configuration of TM to determine a choice of a certain element of a set relevantly (i.e. a nonempty tuple of instructions for that HV).

*Statement 3.3*. Any Turing machine needs an infinite time for FRC (for the infinite set of steps, each of which corresponds to one configuration); symbolically: $FRC \rightarrow (t_{TM} = \infty)$.

The theorem intended to be proved and its corollary follow from *Statement 3.3*. Indeed, the described logical pathway imply the direct implication from its starting point to its end point: $QP \rightarrow (t_{TM} = \infty)$; or by words in English: QP implies infinite time for any Turing machine to resolve it; that is the *Theorem*. One can consider the above sketch of the proof in detail bellow:

*Statement 1*. $QP \rightarrow HV_{K\&S}$.

Indeed, the quantum superposition being a necessary condition for QP implies in turn the theorem of Kochen and Specker, and a necessary of which is their definition of "hidden variables". Thus, QP implies the Kochen - Specker hidden variables.

The present paper introduces the following definition of "hidden variables ($HV_{FC}$)":

$\forall M, \forall x, \exists n, \exists p_1, \exists p_2, \dots p_n: [(x \in M) \leftrightarrow x = M_1 \cap M_2 \dots M_n] \wedge [(M_1 \leftrightarrow p_1), (M_2 \leftrightarrow p_2), \dots, (M_n \leftrightarrow p_n)]$

Here, $M$ is an arbitrary set, $x$ is an element which belongs to $M$, $n$ is a certain natural number; $M_1, M_2, \dots M_n$ are subsets of $M$; $p_1, p_2, \dots p_n$ are the characteristic properties of $M_1, M_2, \dots M_n$ respectively. Then, $p_1, p_2, \dots, p_n$ are called "hidden variables ($HV_{FC}$)" determining the choice:

$$Ch: \left(M \xrightarrow{Ch} x\right) \leftrightarrow Ch = p_1 \wedge p_2 \wedge \dots \wedge p_n = M_1 \cap M_2 \cap \dots \cap M_n$$

Then: $Ch = Ch_1, Ch_2, \dots, Ch_n$ after one means $Ch_i = (M_i \leftrightarrow p_i)$, and the well-ordering notated by the index $i$ means the class of equivalence of the ordinal number $n$ and thus, the cardinal number $n$ both finite and the same natural number $n$. Informally, the concept of hidden variables so defined means that the unambiguous choice of a single certain element of a certain set can be decomposed to an equivalent *finite* set of independent choices.

---

[4] The term "well-ordering principle" is used in some papers in the same meaning. However, "well-ordering principle" is used in two different meanings: (1) as referring to arithmetic and equivalent to the axiom of induction; (2) as referring to set theory and equivalent to the axiom of choice. The axiom of choice and the axiom of induction are absolutely different and referring to two absolutely different theories: set theory and arithmetic respectively. The term "well-ordering principle" implies the above inadmissible ambiguity. The other possible term "well-ordering theorem" originates form history and the tradition: Zermelo (1904) proved it as a theorem implied from the axiom of choice. Whitehead and Russell demonstrated the proof of the converse statement about ten years later, in the beginning of the third volume of *Principia Mathematica*. However, the term "theorem" does not express the equivalence of both statements. Those considerations underlie the term "well-ordering postulate" coined here.

In other words, $Ch$ is the characteristic property of the set $\{x\}$ consisting of a single element $x$. So defined, $HV_{FC}$ are also a finite set of propositions, each of which determines unambiguously a mapping of $M$ into the set of its subsets once $M$ is given, and each proposition chooses one single subset of its. Any of those subsets whether infinite or finite can be considered furthermore as the definition area of one among "$n$" variables $v_1, v_2 \ldots v_n$ of the functions of choice defined on $M$ and determining unambiguously just $x$:

$$\{M \xrightarrow{Ch(v_1, v_2, \ldots, v_n)} x\} \leftrightarrow \{M \xrightarrow{Ch_1, Ch_2, \ldots, Ch_n} x\}$$

The essence of that definition of $HV_{FC}$ is that (1) each hidden variable is an independent choice, and (2) both hidden variables and independent choices are the same *finite* set, each element of which can be interpreted as a hidden variable or an independent choice.

One can admit a few, partly intersected cases of further abstraction: (1) one can prove only the existence of a finite set of HV, but not determine them explicitly; (2) one can define explicitly an infinite set of HV (i.e. the set of HV can be defined by a finite set of characteristic properties); (3) one can prove the existence of an infinite set of HV, but not determine them explicitly; (4) one can postulate the existence of a function of choice for any element of any choice: properly the axiom of choice. It does not imply the existence of any relevant set of its variables determining the chosen element unambiguously or the element chosen once can be chosen again unambiguously. FRC means just the last case. The definition of HV as above can be granted as the initial case, "(0)". If one need elucidate the logical relation between (4) and (0), the cases (1), (2), and (3) should be represented by means of (0) and (4). The axiom of choice (and FRC because of it, in definition) is not referred to any set of hidden variables. Then, the problem is whether one can equate no reference to any set of hidden variables (or the slogan "ho hidden variables" as representing that "no reference" case) as in (4) to some infinite HV set whether known as in (2) or unknown as (3)

Right the equivalence of the axiom of choice and the well-ordering postulate implies the positive answer as the solution of the problem: no HV is equivalent to an infinite set of HV. Indeed, FRC is equivalent to the axiom of choice in definition. Then, the axiom of choice is equivalent to the well-ordering postulate, i.e. to a well-ordered infinite[5] set of chosen elements, each of which implies a nonempty[6], at least finite set of HV, which in turns implies for all HV corresponding to the axiom of choice to be infinite necessarily.

Thus, the equivalence of the axiom of choice and the well-ordering postulate implies both identifications: (1) DC is equivalent to a finite set of HV; (2) FRC is equivalent to an infinite set of HV; (3) if any set is either finite or infinite, any choice is either DC (including HDC) or FRC, and RC can be distributed exhaustively into two disjunctive classes includable either in HDC (and thus in DC) or FRC.

Those considerations suggest

*Statement 2.* $HV_{K\&S} \leftrightarrow HV_{FC}$.

It consists of the following two statements:

Statement 2.1. $Ch_q \rightarrow FRC$. That is: the choice between states of quantum superposition is fundamentally random. The axiom of choice is necessary for it to be accomplished. The determination of the "non-P and NP problems" needs the case of a finite number of quantum states in superposition for a "guessed" solution to be checkable for a polynomial time as an item in a certain n-tuple:

Quantum superposition of a finite number of states involves a continuum (i.e. an actual infinite set) of links between the elements of a finite set of states. Thus, the choice refers to an actual *infinite* set (that of links), and the check of a chosen element to a *finite* set (that of states).

Furthermore, $(Ch_q \rightarrow FRC) \rightarrow (\neg FRC \rightarrow \neg Ch_q) \rightarrow (HV_{FC} \rightarrow HV_{K\&S})$

---

[5] That set can be accepted to be infinite as any finite set can be well-ordered constructively.

[6] An empty set of characteristic properties cannot define any choice at all.

*Statement 2.2.* $HV_{K\&S} \rightarrow HV_{FC}$

Only Statement 2.1 is necessary for the proof that the class of "non-P and NP problems" is not empty.

Both *Statement* 2.1 and *Statement* 2.2 are necessary for the definition of the class of "non-P, but NP" problems in two independent, but equivalent ways: (1) directly, by quantum superposition: as any quantum choice among any finite set of disjunctive alternatives (being in superposition); (2) indirectly, by the concept of hidden variables: as the absence of any hidden variables able to determine any quantum choice finitely.

One can consider the proof of each of both statements independently:

*Statement 2.1.* $Ch_q \rightarrow FRC$

Proof:

Let $A$ be any observable therefore satisfying the equation (1) in the paper of Kochen and Specker (1967: 61). Different values $A_i$ of $A$ are measured by a series of measurement $M_i$. One considers the set $\{A_i\}$ defined by the property any element $A_i$ to be a measured value of the observable $A$. The well-ordering postulate applied to $\{A_i\}$ creates a well-ordering of $\{A_i\}$: $WO_A = \{A_{i=1,2,\dots\infty}\}$. Then $WO_A$ implies that the equation (4) in the paper of Kochen and Specker (1967: 64) is satisfied. Indeed, that equation (4) is:

$$P_{g(A)\psi}(U) = P_{A\psi}g^{-1}\big((U)\big)$$

Here $A$ is an observable, $\psi$ is a quantum state represented by its wave function unambiguously, $g$ is a Borel function of the set of real numbers $R$ onto itself, and $U$ is a certain subset of $R$ which is measurable by a probability measure $P_{A\psi}$ assignable to each observable A and each quantum state $\psi$ by a certain function $P$. Informally, this means that a necessary condition for HV is the inverse function of any statistical distribution assignable to each observable and each quantum state to be a statistical distribution as well.

Interpreting that equation (4) to the present proof, it would require for the inverse function to the function of a series of measurements thus representing any quantum superposition of the observable $A$ as an unambiguously corresponding well-ordering $WO_A$ to be again a well-orderingl. However, the latter contradicts to the definition of "quantum superposition".

Indeed, the quantum superposition implies for any two measured values of any observable, the superposition of their corresponding states (i.e. their sum with arbitrary complex coefficients) to be a valid state. Thus, the inverse mapping cannot be a Borel function as it is not any function at all: to those two measured values correspond a continuum of values in the inverse mapping at issue.

That is: $A$ would be a hidden variable if $WO_A$ existed before the series of measurements. However, this contradicts the Kochen-Specker theorem. Consequently, $WO_A$ appears after the series of measurements. One need equate the quantum superposition before the series of measurements and $WO_A$ after it. This implies that the series of measurements involves necessarily the well-ordering postulate[7]. The well-ordering postulate is equivalent to the axiom of choice. Thus, any measurement is FRC, and *Statement* 2 is proved.

The proof can be justified furthermore by the following physical interpretation. Quantum electrodynamics relies on the mutual consistency of quantum mechanics and special relativity. So, the quantity of time, which is not an observable in the nonrelativistic quantum mechanics, has to be equated to all other observables in quantum electrodynamics. This can be accomplished in any of the following two ways:

1. The observable of time to be allowed in quantum electrodynamics.

---

[7] One can notice that the series of measurements is well-ordered by itself: the first measurement, the second measurement, and so on. The well-ordering postulate is necessary because the actual infinity of all measurements of the observable $A$ is involved.

2. On admits that any observable (most frequently, space position) to be only a parameter, or "label" naming and distinguishing the quantum states, particularly those in any quantum superposition.

Each of the above two options together with the Kochen-Specker theorem implies *Statement 2*. Thus, the validity of the Kochen-Specker theorem in quantum electrodynamics implies *Statement 2*. More exactly, The Kochen-Specker theorem and the consistency of quantum mechanics and special relativity[8] implies *Statement 2*.

However, *Statement* 2 can be released from its interpretation in terms of quantum mechanics and thus generalized if one considers an arbitrary class (for the quantum superposition of the observable $A$) defined by any consistent property and then, that class is mapped one-to-one (for the measurement) into a certain set (for the set $\{A_i\}$ of all measured values of $A$). The above proof refers to an interpretation, properly a quantum interpretation, of the same isomorphic structure in general as here.

*Statement* 2.2. $HV_{K\&S} \rightarrow HV_{FC}$

Proof:

The necessary condition for $HV_{K\&S}$ to exist (Kochen, Specker 1967: 66) will be utilized in order to be demonstrated that it implies $HV_{FC}$ and *Statement 2.2* to be proved as follows.

One can demonstrate that *Theorem 1* in the paper of Kochen and Specker (1967: 70) about the incommensurability involved by a finite partial Boolean algebra follows from one of the most ancient mathematical proofs about the incommensurability of the diagonal of any square to its sides (i.e. the proof that $\sqrt{2}$ is an irrational number). To be shown that, one need define the separable complex Hilbert space as a generalization of arithmetic, and then, any partial finite Boolean algebra as a generalization of the arithmetic conception of rational number: the Pythagorean proof about the incommensurability of the diagonal of the square can be generalized as that *Theorem 1* at issue.

In turn, that *Theorem 1* (Kochen, Specker 1967:70) together with the necessary condition for $HV_{K\&S}$ to exist (cited above) is involved in a *modus tollens* deduction to be proved the absence of those $HV_{K\&S}$ in their paper. Instead of that, one generalizes that necessary condition in the way hinted above for proving *Statement 2*.2 It follows immediately from the necessary condition generalized so. That plan for proving *Statement 2.2* is to be explicated in detail:

Its core is the generalization of arithmetic meant for certainty by the Peano axioms as usual where any natural number $n$ is considered as a class of equivalence of all sets consisting of $n$ elements, or with a cardinal number $n$ shared by all set in the same class. The approach by a class of equivalence to be define any natural number $n$ will be conserved, but what is equivalent will be changed: qubits rather than sets:

A unit is the class of equivalence of all possible values of a qubit, or an "empty" qubit as an empty cell, in which no value is recorded yet following the scheme, by which "Turing machine" is defined. So, if one considers the class of equivalence of all wave functions consisting of $n$ successive qubits, the natural number $n$ is defined[9].

A qubit is defined as the superposition of any two orthogonal states or subspaces of the separable complex Hilbert space, $\langle 0|1 \rangle$ by complex coefficients $\alpha, \beta$ so that: $[\alpha]^2 + [\beta]^2 = 1$. That is, "qubit" $Q \stackrel{\text{def}}{=} \alpha|0\rangle + \beta|1\rangle$, and any pair $\alpha, \beta$ of complex numbers under the above condition are a value of qubit. As any two successive axis of the separable complex Hilbert space, $e^{in}, e^{i(n+1)}$ are a particular case of two orthogonal states of the separable complex Hilbert space, any element of it, interpreted in quantum mechanics as a wave function, can be equivalently represented as a series of qubits.

---

[8] The same consistency of quantum mechanics and special relativity is a premise in the so-called freewill theorems (Conway, Kochen 2006; 2009). Furthermore, their meaning is similar to that of *Statement 2.1*. They can be considered as relative to *Statement 2*.

[9] This is a "natural arithmetic" isomorphic to that defined standardly, but extracted from the physical nature itself by means of quantum mechanics rather than from human activity and thus, more or less conventionally.

Indeed, a wave function is standardly represented as the vector: $C_1, C_2, \ldots, C_n, C_{n+1}, \ldots$, where $\{C_i\}$ is a convergent series of complex coefficients. It can be transformed unambiguously in the vector: $Q_1, Q_2, \ldots, Q_n, Q_{n+1}, \ldots$ where

$$Q_i = (\alpha_i, \beta_i) = \frac{C_i}{\sqrt{(C_i)^2 + (C_{i+1})^2}}, \frac{C_{i+1}}{\sqrt{(C_i)^2 + (C_{i+1})^2}}$$

"Qubit" can be considered as the generalization of "bit", referring to infinite sets, and thus to set theory properly. Indeed, "bit" referring only to finite sets is a notion rather arithmetical. It means an elementary choice between two equally probable alternatives, and serves as a measure to represent the quantity of those elementary choices (the quantity of "information") necessary to be chosen any single element from any finite set. It is useless for the quantity of choices necessary to be chosen an element from an infinite set.

A new measure for a new quantity is necessary to represent the choice from an infinite set:

*Statement A*: The concept of qubit $Q$ is necessary and sufficient to represent the choice $Ch_\infty$ of a single element from any infinite set.

To be a single choice from any infinite set possible always, the axiom of choice is necessary and sufficient condition. Consequently it a premise granted for the statement.

*Sufficiency $Q \to Ch_\infty$*:

Any qubit is a continuum and thus an infinite set. Any certain value of it represents a choice of a single element from an infinite set, i.e. $Ch_\infty$; consequently, $Q \to Ch_\infty$.

*Necessity $Ch_\infty \to Q$*:

The axiom of choice implies the Skolem "relativity of the concept set" because it implies the necessary existence of some bijective mapping between any two infinite sets. Thus, some bijection exists between continuum (and particularly, any qubit) and any infinite set. Consequently any choice from the latter can be represented equivalently as some choice from the former: $Ch_\infty \to Q$.

The concept of qubit is introduced by quantum mechanics. Nonetheless, it has a much more general meaning referring to the generalization of "information" to set theory:

A complex quantity named quantum information $I_q$ and measured in qubits can be introduced. Any wave function is a certain value of $I_q$. The free variable of $I_q$ is isomorphic to the separable complex Hilbert space.

*Statement B*: The quantity of quantum information $I_q$ is necessary and sufficient to represent any choice from any infinite set $Ch_\infty$. .

*Necessity $Ch_\infty \to I_q$.* Any choice from any infinite set can be defined as a class of elementary choices, qubits, which, furthermore, is a *set* of the same elementary choices, qubits. This is true because "any choice from any infinite set" determines a certain element of the set of all subsets of that infinite set at issue. Thus, the "class of elementary choices" refers to a certain set unambiguously. Then, let one consider that "set of elementary choices", which is an arbitrary set. It can be represented as a well-ordered series of qubits in virtue of the axiom of choice. That well-ordered series of qubits in turn is equivalent to a wave function, and thus, to a certain value of quantum information.

*Sufficiency $I_q \to C_\infty$.* . Indeed, it is obvious: $I_q \to Q \to C_\infty$

Then, the paper of Kochen and Specker (1967) referring only to any elements of the separable complex Hilbert space in the final analysis (and thus to all of them) can be interpreted thoroughly arithmetically one the above definition of arithmetic is introduced. Indeed, for example, the quantum superposition of two disjunctive possible states is an empty qubit, and thus, a unit in the so defined quantum arithmetic; and the crucial concept of finite partial Boolean algebra would include all rational numbers in the quantum arithmetic and would imply the necessary existence of irrational numbers, i.e. *Theorem 1* (Kochen, Specker 1967: 70) as an interpretation of the Pythagorean discovery about two millennia and a half ago.

One need consider what $HV_{K\&S}$ and their necessary condition will represent after that transformation from the separable complex Hilbert space to arithmetic by classes of equivalence, i.e. from qubits to arithmetical units: any element of any partial Boolean algebra of observables in quantum mechanics will correspond to an ordered pair of natural numbers and thus to their fraction, i.e. to a certain rational number. Indeed, one has to count the number of qubits, in which any certain value is available (or "recorded"). Thus, any wave function will be mapped in a certain natural number. One is to emphasize expressively that "all natural numbers", i.e. the concept of *actual infinity is not involved in relation to the number of axis* (equal to the number of qubits, in which certain value is available) of the separable complex Hilbert space after its transformation by classes of equivalence into arithmetic[10]. Thus, any wave function associated with one or more observables[11] will be mapped as a certain natural number: and *each natural number is finite*, which follows immediately from the axiom of induction in Peano arithmetic.

Indeed, the set of rational numbers represented as ordered pairs of integers satisfies the definition of finite partial Boolean algebra:

For example, one generates a partial Boolean algebra by the separable "$Z_2$"[12] Hilbert space. Each vector in it is a binary positional notation of just one certain natural number. Then, the Kochen – Specker commensurability ⚥ can coincide with the usual arithmetical commensurability defined by the availability of a common divisor.

One can remove the mediation of the binary positional system by admitting the degenerated field $Z_1$ consisting of a single element and where addition and multiplication coincide necessarily. The corresponding vectors in the separable $Z_1$ Hilbert space are interpretable directly as natural numbers or respectively, as their ordinals, i.e. as results of the usual counting of the elements of any finite set. That finite set can consist of qubits, which corresponds to the definition of an arithmetical units as the class of equivalence of all values of a qubit.

Nonetheless whether $Z_2$ or $Z_1$ is used for the transition to arithmetic, the standard arithmetical commensurability satisfies the properties $1 - 4$ (Kochen, Specker 1967: 64) postulated for the commensurability ⚥ necessary for any partial algebra to be defined. Then, the ordered pairs of natural number (i.e. the rational numbers) constitutes a partial finite Boolean algebra.

A minimal geometrical structure (e.g. the plain, the two-dimensional Euclidean space, as in the original Pythagorean proof of incommensurability) is necessary to be added to the arithmetic for demonstrating incommensurability. In fact, the geometrical structure only exemplifies the concept of actual infinity, i.e. any actually infinite set. Even being countable, it is sufficient to generate incommensurability for any finite partial Boolean algebra. In turn, the partial Boolean algebra only exemplifies any arithmetical structure sufficient to generate incommensurability being combined with the "actual infinity" of set theory[13].

---

[10] Even more, it cannot be involved in any consistent way. However, this statement need not be proved because is not necessary for the present proof.

[11] The case of more than one observable is discussed in the paper of Kochen—Specker (1967: 74-75). However, this case need not be discussed here respectively for any wave function associated with whether one or more observable is mapped as a certain natural number necessarily.

[12] The finite field $Z_2$ consisting of two elements is utilized as in the paper of Kochen and Specker for partial Boolean algebras as in *Appendix 1* of present paper for the equivalent redefinition of Turing machine on the separable $Z_2$ Hilbert space. Here, it will used also for the definition of "partial Boolean algebra" in arithmetic. The link, historically and traditionally, is directed oppositely: the arithmetical commensurability (respectively, incommensurability discovered by the Pythagorean school) is generalized by the concepts of "partial algebra" and "partial Boolean algebra" in their paper for the investigation of the problem of hidden variables in quantum mechanics. However, the present paper demonstrates that the arithmetical incommensurability is sufficient for the absence of hidden variables in quantum mechanics. That sufficiency is explicated by considering the separable complex Hilbert space of quantum mechanics as a generalization of arithmetic, or respectively, by considering arithmetic by classes of equivalents in that Hilbert space.

Furthermore, if one interprets those ordered pairs of natural numbers as orthogonal vectors in the plain or in any other finitely dimensional Euclidean space, the length of the "diagonal" in the plain or a certain length in any other finitely dimensional Euclidean space (corresponding to the "finite partial Boolean algebra D" in *Theorem 1* in the paper of Kochen and Specker (1967: 70) will be incommensurable with the natural numbers, i.e. it is an irrational number.

In conclusion, any pair of wave functions being an element of any partial Boolean algebra of observables are mapped as a certain rational number after transition from the separable complex Hilbert space to arithmetic by classes of equivalence of qubits.

The hidden variables $HV_{K\&S}$ are defined as observables. A certain natural number $n$ thus finite corresponds to any observable represented by "$n$" qubits after the transformation in arithmetic by classes of equivalence. Each of those "$n$" qubits can be consider as an independent choice, totally a finite set consisting of "$n$" elements as well. Thus, a finite set of $HV_{FC}$ is determined unambiguously. *Statement 2.2* is proved.

The essence of that simple proof of *Statement 2.2* relies on the disjunctive distinguishability of the arithmetical finiteness from the set theory actual infinity in the concept of the separable complex Hilbert space in the final analysis. Its redefinition by means of qubits as well as the quantum definition of Peano arithmetic visualizes that distinguishability: actual infinity is only "within" the qubits, which as classes of equivalence, i.e. "outside" of them can be considered only arithmetically. Furthermore, the same approach visualizes the implicit arithmetical essence of the Kochen-Specker theorem as generalizing and thus corresponding to the ancient proof about the existence of incommensurability A statement sufficient for the proof is that any single $Ch_q$ between a finite set of states implies an infinite set of hidden variables. *Statements 1 and 2* being more general imply it. This has to be translated in terms of "Turing machine" according to *Official Problem Description*. First of all, "hidden variable" needs that translation:

Any hidden variable is equivalent to a set whether finite or infinite and to its characteristic property finite always. Only the latter can be processed by a Turing machine in general. The characteristic property can be meant in the program only, unlike the elements of a certain set, which are transformed into those of another set by the program. Indeed, the definition of Turing machine requires for it to process only finite sets by finite programs in order to be able to finish in any finite time for the time of a single step is neither zero nor infinitesimally small. Thus, any Turing machine can mean any infinite set only by means of its characteristic property as a finite tuple of instructions. Then, any hidden variable being equivalent to both finite and infinite sets in general has to refer to a certain finite tuple of instructions unambiguously.

Anyway, if some hidden variable corresponds to a finite set, it can be represented absolutely in the tape of Turing machine, particularly in the input string, as a certain substring corresponding one-to-one to that hidden variable. This corresponds to defining a set by the complete enumeration of its elements not needing any characteristic property. Thus, the corresponding tuple of instructions can be empty. If that is the case, only an infinite simultaneous substring can represent unambiguously an infinite set of hidden variables, each of which is represented in turn by a finite set on the tape (as far as the corresponding tuple of instructions is empty an no movement of tape is possible). This implies: (1) either the input string to be infinite, which the definition of Turing machine does not admit; (2) or the length of the tape to include an infinite set of cells simultaneously. The latter implies an infinite time for processing because that infinite set cells needs infinite set of steps (configurations) as it to be obtained from the finite input string as it to be processed.

---

[13] That kind of incommensurability (i.e. the incommensurability of arithmetical finiteness and set-theory actual infinite) is inferred as "incompleteness" in the famous paper of Kurt Gödel published in 1931. It is not cited in the present paper for it is mentioned only as an illustration not being used as a premise in the deduction, furthermore being well-known.

Furthermore, that finite tuple cannot be empty for no set is defined by a zero tuple for there is no characteristic property. Consequently any hidden variable corresponds unambiguously to a tuple containing at least one instructions.

Briefly, any infinite set of hidden variables needs an infinite set of finite tuples of instructions (roughly speaking, an "infinite program") or an infinite set of cells simultaneously as a substring of the tape (roughly speaking, an "infinite tape"). Each of them implies an infinite set of steps of Turing machine and thus an infinite time of work.

This consideration can be summarized as a few statements:

*Statement 3*. The finite set of unambiguous criteria for an element to be chosen from its set is equivalent to a determined choice and to the negation of a fundamentally random choice: $HV_{FC} \leftrightarrow DC \leftrightarrow \neg(FRC)$.

*Statement 3*.1. Any Turing machine can choose a certain element of a set only by DC (i.e. by a finite tuple of instructions).

*Statement 3.2*. Any parameter (i.e. a certain value of a certain HV) needs at least one configuration of TM to determine a choice of a certain element of a set relevantly (i.e. a nonempty tuple of instructions for that HV).

*Statement 3.3*. Any Turing machine needs an infinite time for FRC (for the infinite set of steps, each of which corresponds to one configuration).

The last statement implies immediately the theorem which had to be proven as well as its corollary, which will be used further (because FRC is equivalent to QP according to *Statements 1 and 2* above):

*Theorem*: No Turing machine can resolve QP for any finite time.

*Corollary*: QP belongs to the class of non-P problems.

*IV A description of both problem and solution in the notations of the "Official Problem Description" of CMI:*

The Turing machine trying to resolve the problem deterministically and denoted by "TM" is a Turing machine as it is defined in the appendix of the Official Problem Description. Let its input alphabet consist of two symbols: $\Sigma = ("0", "1")$. Let $\Sigma^*$ be the set of finite strings written by the alphabet "$\Sigma$" consisting of two elements. Let $\omega$ be an element of $\Sigma^*$ which is the input of $R_n$. As the computation of TM never halts, TM does not accept the string $\omega$. Thus, the calculation of the input string $\omega$ by "TM" does not belong to "P".

Nonetheless, it belongs to "NP". Indeed:

The checking Turing machine ("CTM") only tests deterministically a hypothetical solution X obtained non-deterministically somehow, e.g. "guessed". Let its input alphabet consist also of the same two symbols $\Sigma_1 = ("0", "1")$. The problem of CTM is whether X belongs to an "n-tuple" consisting of $R_n$ or not. The checking relation is defined so: $CR = R_n \times X \subset \Sigma^* \times \Sigma_1^*$. Thus, $CR$ consists of "$n$" elements as well as the associate language $L_{CR}$. Consequently, $L_{CR} \in P$, and $CR$ "is polynomial time":

Thus, the language of "TM" belongs to NP since $|X|$ is finite, and then:

$\forall R \in R_n, \exists k \in \mathbb{N}: \{|X| \leq |R|^k$ and $CR(R, X)$ is polynomial-time$\}$.

Obviously, the restriction for both alphabets $\Sigma$ and $\Sigma_1$ to consist of only two symbols is incidental: it originates from the practical realization of contemporary computer. Any of both alphabets can consists of any number of symbols, satisfying the "Official Problem Description"[14], any finite natural number such that $|\Sigma|, |\Sigma_1| \geq 2$.

---

[14] "It is easy to see that the answer is independent of the size of the alphabet $\Sigma$ (we assume $|\Sigma| \geq 2$), since strings over an alphabet of any fixed size can be efficiently coded by strings over a binary alphabet" (p. 2).

The assumption is that X is an arbitrary rational (or real) number represented with a certain accuracy as a finite string. However, the formulation of the problem imposes the requirement for X to be a certain element of the n-tuple $R_n$. Obviously, the solution above includes that restriction as a particular case.

A class of problems is proved to be NP, but not P. Informally, that class includes all problems about the *fundamentally random choice* of an alternative among a *finite* set of possible alternatives. The Kochen – Specker theorem allows for a formal definition of "fundamentally random choice", which can be discussed further in terms of Turing machine so that it can be proved to be undecidable problem for any Turing machine needing "infinite time" to resolve it and thus, being a non-P problem. Nonetheless, this is a NP problem since a finite set of possible solutions is necessary to be checked.

*V Two conjectures for further research:*

Intuitively, the following consideration seems to be convincing. Any non-P problem needs an infinite time to be resolved because any finite time is polynomial in the final analysis. This means that the following *first conjecture* would be granted: *there exists an algorithm of Turing machine able to resolve any problem for a polynomial time if there exists an algorithm of Turing machine able to resolve the same problem for any finite time*. On the contrary, any NP problems needs a certain finite and thus, polynomial time to be resolved. Consequently, the "P vs NP" should be isomorphic to a one-to-one mapping of an infinite set into a finite one: at first glance, a mistake in definition as far as the corresponding set seems to be empty.

However, at least one class of solutions was demonstrated above and that set is not empty.

A *second conjecture* is that *the class at issue includes all possible solutions of the "P vs NP"*.

One need prove that only a fundamentally random choice is able to map an infinite set into a finite set one-to-one, furthermore. Outlines of that proof might be:

The *finite* set, into which any infinite set is able to be mapped one-to-one is granted as a Dedekind finite set, i.e. there does not exist any one-to-one mapping of that single infinite set into any set being a "Dedekind finite set".

In order not to exist any one-to-one mapping, it is sufficient not to exist any mapping of that kind at all (i.e. including any one-to-one mapping). The latter is satisfied if the infinite set at issue is "mapped" into a set of finite sets as it is not properly a true mapping so that only one single certain set to correspond to the infinite set. The last "set of finite sets" in turn can be as (1) finite as (2) infinite.

Though the case (2) by itself is very interesting, one can ignore it in the present context because it implies "non-NP".

If (1) is the case, one may exclude any "hidden variables"[15], which would determine the choice of a certain finite set as the image of the infinite set implicitly. Indeed, if the latter could be the case, one might complement the defective "mapping" to a normal mapping by means of relevant values of those hidden variables. However, the normal mapping of an infinite set into a (non-Dedekind) finite set is impossible, thus, the "hidden variables" are also impossible in virtue of *modus tollens*.

If no hidden variable can exist, this is equivalent for the choice of a finite set as corresponding to the infinite set to be fundamentally random, at least intuitively convincingly.

The method involving the Dedekind finiteness for proving the second conjecture need be "doubled" for proving the first conjecture. One utilizes two probabilistic mappings (each implying a different statistic distribution) of the same countable set into two different finite sets rather than only a single one as for proving the second conjecture. This is what is meant as the method to be "doubled".

Then, the one-to-one mapping of the countable set into the first finite set implies the reverse bijective mapping. Further, a bijective mapping composed by the latter one (as the first one in the composed mapping) and the other mapping of the countable set in the second finite set (as the second one in the composed mapping exists necessarily. It can be defined to be a "Dedekind bijective mapping" between any two finite sets consisting of different numbers of elements.

---

[15] The introduced term coincides with the analogical one in quantum mechanics purposely.

Then, the same kind of "Dedekind bijective mapping" is able to equate any algorithm consisting of any finite number steps (e.g. an algorithm ending for any finite time) to any other of the same kind (e.g. an algorithm ending for any polynomial time). The proof by "Dedekind bijective mapping" would prove only "pure existence" for the latter algorithm because of the only statistical link between the former and the latter algorithm.

*Appendix 1*: Redefinition of Turing machine in terms of the separable "$Z_2$"[16] Hilbert space

The redefinition extends the arithmetical definition of Turing machine (without any reference to "actual infinity") to an equivalent definition, to which the "infinite" calculations of quantum computer can make sense. This is not necessary for the formal proof of the "P vs NP problem", but for the meaning of the class "non-P, but NP" to be understood.

The separable "$Z_2$" Hilbert space is defined on the field of two elements, e.g. so that the axiom of fil The redefinition consists in the replacement of one interpretation of a formal structure by another and equivalent interpretation of the same formal structure. Thus, the redefinition represents the successive equivalent replacement of the terms that is necessary to be substituted in the "Appendix" of the "Official Problem Description" by the terms relevant to the separable "$Z_2$" Hilbert space. Anyway, what need not be replaced is repeated without change to be conserved the context. What is substituted or added is underlined in order to be clear what is changed:

A Turing machine M consists of a finite state control (i.e., a finite program) attached to a read/write head moving on an infinite tape. <u>The separable "$Z_2$" Hilbert space is the tape T with a specified input alphabet $\Sigma$ consisting of the two elements of $Z_2$, which is a true subset of the alphabet $\Gamma$ consisting of three symbols: the two elements of $Z_2$ and b meaning any axis of T.</u> At each step in a computation, M is in some state q in a specified finite set Q of possible states. Initially, <u>an infinite subspace of T is chosen by a finite (complementing) vector $V_0$ of T (thus, "written" on adjacent axes of T)</u>, the head scans the left-most symbol of the <u>vector $V_0$,</u> and M is in the initial state $q_0$. At each step M is in some state q and the head is scanning a tape <u>axis</u> containing some <u>value $s$</u>, and the action performed depends on the pair (q, s) and is specified by the machine's transition function (or program). The action consists of <u>the choice of a certain value</u> on the scanned <u>axis</u>, moving the head left or right one <u>axis</u>, and assuming a new state.

Formally, a Turing machine M is a tuple $(\Sigma, \Gamma, Q, \delta)$, where $\Sigma, \Gamma, Q$ are finite nonempty sets with $\Sigma \subseteq \Gamma$ and $b \in \Gamma - \Sigma$. <u>Here, "$\Sigma$" is interpreted as the two elements of $Z_2$, "$\Gamma$" as any axis of T, and "b" means the "empty" axis, on which no value (i.e. no element of $Z_2$) is chosen.</u> The state set Q contains three special states $q_0, q_{accept}, q_{reject}$. The transition function $\delta$ satisfies:

$$\delta : \left(Q - \{q_{accept}, q_{reject}\}\right) \times \Gamma \to Q \times \Gamma \times \{-1, 1\}$$

If $\delta(q, s) = (q', s', h)$, the interpretation is that, if M is in state q scanning the <u>value</u> s, then q' is the new state, s' is the <u>value changed</u>, and the tape head moves left or right one square depending on whether h is −1 or 1. <u>This means that the transition function $\delta(q,s)$ can choose exactly one option according to the program among a few: either (1) q'=q+h (three disjunctive options), and "s" does not matter; or (2) s=s' (two disjunctive options: "s" either conserves or changes), and "h" does no matter. "q" is interpreted as a certain finite dimension of T; "q-1" means a certain vector, and "q+1" either a certain vector or a *relevant* "blank" subspace of T; both "s" and "s'" can be equal to any of the two values of $Z_2$.</u>

We assume that the sets Q and $\Gamma$ are disjoint. <u>This means that $\Gamma$ is restricted to refer only to the infinite subspace of T unambiguously determined by any finite Q.</u>

A configuration of $M$ is a string $xqy$ with $x, y \in \Gamma^*$, $y$ not the empty string, and $q \in Q$. <u>The language $\Gamma^*$ is defined on T as follows: all ordered pairs of (i) a finite vector of T (corresponding to the alphabet $\Sigma$) and (ii) a finite subspace T (corresponding to the symbol b = $\Gamma - \Sigma$).</u>

The interpretation of the configuration $xqy$ is that M is in state $q$ with <u>$xy \in \Gamma^*$</u>, with its head scanning the left-most symbol of $y$ <u>where $x$ is a finite vector of T, and $y \in \Gamma^*$.</u>

If $C$ and $C'$ are configurations, then $C \xrightarrow{M} C'$ if $C = xqsy$ and $\delta(q, s) = \delta(q', s', h)$ and one of the following holds:

"$C' = xs'q'y$ and $h = 1$ and $y$ is nonempty.

---

$C' = xs'q'b$ and $h = 1$ and $y$ is empty.

$C' = x'q'as'y$ and $h = -1$ and $x = x'a$ for some $a \in \Gamma$.

$C' = q'bs'y$ and $h = -1$ and $x$ is empty.

A configuration $xqy$ is halting if $q \in \{q_{accept}, q_{reject}\}$. Note that for each nonhalting configuration $C$ there is a unique configuration $C'$ such that $C \xrightarrow{M} C'$.

The computation of M on input $\omega \in \Sigma^*$ (where is $\Sigma^*$ is a language written by the alphabet $\Sigma$) is the unique sequence $C_0, C_1, \dots$ of configurations such that $C_0 = q_0\omega$ (or $C_0 = q_0b$ if $\omega$ is empty) and $C_i \xrightarrow{M} C_{i+1}$ for each $i$ with $C_{i+1}$ in the computation, and either the sequence is infinite or it ends in a halting configuration. If the computation is finite, then the number of steps is one less than the number of configurations; otherwise the number of steps is infinite. We say that M accepts $\omega$ iff the computation is finite and the final configuration contains the state $q_{accept}$.

Informally, any Turing machine finds a certain result ($q = q_{accept}$) or not ($q = q_{reject}$) starting from a certain input $\omega$ for a finite number of configurations or not. Those three cases are disjunctively. In fact, the definition by Hilbert space T as above is a generalization as far as it admits to find a result ($q = q_{accept}$) not for a finite number of configurations. The generalization (unlike a properly arithmetical definition of Turing machine) needs "actual infinity" for the transfinite calculations to make sense:

New and new dimensions of T are added to the dimension of the finite input $\omega$ according to the program unambiguously until a result be obtained ($q = q_{accept}$) or not ($q = q_{reject}$). A calculation being defined by Hilbert space T can continue as if "in infinity" where it also can obtain a result ($q = q_{accept}$) or not ($q = q_{reject}$). If any Turing machine makes that calculation, it would need an infinite number of configurations and thus, infinite time. However, if the same calculation is made by a quantum computer, it can finish in a certain (even zero, theoretically) time. Once a certain result of that calculation whether ($q = q_{accept}$) or not ($q = q_{reject}$) is delivered by a quantum computer, it can be checked by a Turing machine for some polynomial time("non-P, but NP) or not ("non-P and non-NP"): the paper demonstrates that the former class is not empty.

*Appendix 2.* Definition of quantum computer as a generalization of Turing machine

The utilized subclass of non-P problems cannot be resolved by any Turing machine for any finite time. Their definition relies on the concept of quantum superposition and involves implicitly "quantum computer" as that calculating machine able to resolve them in some finite time.

Quantum computer is defined bellow as a minimal generalization of Turing machine allowing for an actual infinite set as its alphabet $\Gamma$ consisting of infinite set $\Sigma$ of possible values of any single qubit and their class of equivalence, the same for all qubits, the single symbol "blanc" $b$: the complement $\Gamma - \Sigma$ of $\Sigma$ to $\Gamma$. This implies an infinite set of words admissible in the language $\Sigma^*$ and writable by the infinite alphabet $\Sigma$. The program of quantum computer is also finite as that of Turing machine.

Informally, this means that all bits of the tape of Turing machine are replaced by qubits on the corresponding tape of quantum computer and this is the *single and exhausting difference* between Turing machine and quantum computer. Furthermore, that model is sufficient to describe the work of any real quantum computer as far as it accomplishes a certain finite program utilizing a finite set of qubits. Indeed, if that is the case, the tape of quantum computer is the separable complex Hilbert space. The state set Q of quantum computer refers to some *finite* subspace of the separable complex Hilbert space. Any other state set $Q_\infty$, which refers to some infinite subspace of its, can be equivalently redefined to the former case of finite subspaces. The actual infinity implicitly involved in the definition of a single qubit is sufficient to describe any infinite calculation[17] or at least a very wide class of infinite calculations. That assumption allows for conserving the program of quantum computer as finite as the same or isomorphic to that of Turing machine. Indeed, any infinite configuration of quantum computer can be equivalently described by a certain finite configuration, to which is added a single specific qubit being able to describe exhaustively the difference between meant infinite configuration and the certain finite configuration.

The above redefinition of Turing machine in terms of the separable "$Z_2$" Hilbert space can be considered also as an intermediate stage for defining quantum computer as follows, where $Z_2$ is generalized to the field of complex numbers.

What follows is a new, relevantly generalized repetition of the above redefinition of Turing machine already in terms of the separable *complex* Hilbert space. What is changed in comparison to the redefinition of Turing machine in *Appendix 1* is italicized. It corresponds unambiguously to what is underlined in the previous redefinition: both redefinitions share the same context of the official definition of Turing machine, which is not changed on both cases and being not italicized nor underlined. Anyway, the term "Turing machine" is replaced by "quantum computer" in the following redefinition even in the context, and this only differs the contexts in both cases:

A *quantum computer C* consists of a finite state control (i.e., a finite program) attached to a read/write head moving on an infinite tape. The separable *complex* Hilbert space is the tape T with a specified input alphabet $\Sigma$ consisting of the *infinite set of* elements of *the field of complex numbers,* which is a true subset of the alphabet $\Gamma$ consisting of *an infinite set of* symbols: *the infinite set of elements of the field of complex numbers,* and b meaning any axis of T. At each step in a computation, M is in some state q in a specified finite set Q of possible states. Initially, an infinite subspace of T is chosen by a finite (complementing) vector $V_n$ of T (thus, "written" on adjacent axes of T), the head scans the left-most symbol of the vector $V_0$, and M is in the initial state $q_0$. At each step M is in some state q and the head is scanning a tape axis containing some value $s$, and the action performed depends on the pair (q, s) and is specified by the machine's transition function (or program). The action consists of the choice of a certain value on the scanned axis, moving the head left or right one axis, and assuming a new state.

---

[17] Of course, this is only a conjecture needing a rigorous mathematical proof. Furthermore, this conjecture is equivalent to the assumption that quantum computer is able to resolve *any* problem for a certain *finite* time.

Formally, *a quantum computer C is a tuple* $(\Sigma, \Gamma, Q, \delta)$ *infinite in general, where only Q is a finite nonempty set, and* $\Sigma, \Gamma$ *are infinite sets in general* with $\Sigma \subseteq \Gamma$ and $b \in \Gamma - \Sigma$. Here, "$\Sigma$" is interpreted as the *infinite set of the field of complex numbers,* "$\Gamma$" as any axis of T, and "b" means the "empty" axis, on which no value (i.e. no element of *the field of complex numbers*) is chosen. The state set Q contains three special states $q_0, q_{accept}, q_{reject}$. The transition function $\delta$ satisfies:

$$\delta : \left( Q - \{q_{accept}, q_{reject}\}\right) \times \Gamma \rightarrow Q \times \Gamma \times \{-1, 1\}$$

If $\delta(q, s) = (q', s', h)$, the interpretation is that, if M is in state q scanning the value s, q' is the new state, s' is the value changed, and the tape head moves left or right one square depending on whether h is −1 or 1. This means that the transition function $\delta$(q,s) can choose exactly one option according to the program among a few: either (1) q'=q+h (three disjunctive options), and "s" does not matter; or (2) s=s' (two disjunctive options: "s" either conserves or changes), and "h" does no matter. "q" is interpreted as a certain finite dimension of T; "q-1" means a certain vector, and "q+1" either a certain vector or a *relevant* "blank" subspace of T; both "s" and "s'" can be equal to any of the two values of $Z_2$.

We assume that the sets Q and $\Gamma$ are disjoint. This means that $\Gamma$ is restricted to refer only to the infinite subspace of T unambiguously determined by any finite Q.

A configuration of $M$ is a string $xqy$ with $x, y \in \Gamma^*$, $y$ not the empty string, and $q \in Q$. The language $\Gamma^*$ is defined on T as follows: all ordered pairs of (i) a finite vector of T (corresponding to the alphabet $\Sigma$) and (ii) a finite subspace T (corresponding to the symbol $b = \Gamma - \Sigma$).

The interpretation of the configuration $xqy$ is that M is in state q with $xy \in \Gamma^*$, with its head scanning the left-most symbol of $y$ where $x$ is a finite vector of T, and $y \in \Gamma^*$.

If $C$ and $C'$ are configurations, then $C \xrightarrow{M} C'$ if $C = xqsy$ and $\delta(q, s) = \delta(q', s', h)$ and one of the following holds:

"$C' = xs'q'y$ and $h = 1$ and $y$ is nonempty.
$C' = xs'q'b$ and $h = 1$ and $y$ is empty.
$C' = x'q'as'y$ and $h = -1$ and $x = x'a$ for some $a \in \Gamma$.
$C' = q'bs'y$ and $h = -1$ and $x$ is empty.

A configuration $xqy$ is halting if $q \in \{q_{accept}, q_{reject}\}$. Note that for each nonhalting configuration $C$ there is a unique configuration $C'$ such that $C \xrightarrow{M} C'$.

The computation of M on input $\omega \in \Sigma^*$ (where is $\Sigma^*$ is a language written by the alphabet $\Sigma$) is the unique sequence $C_0, C_1, \ldots$ of configurations such that $C_0 = q_0\omega$ (or $C_0 = q_0 b$ if $\omega$ is empty) and $C_i \xrightarrow{M} C_{i+1}$ for each $i$ with $C_{i+1}$ in the computation, and either the sequence is infinite or it ends in a halting configuration. If the computation is finite, then the number of steps is one less than the number of configurations; otherwise the number of steps is infinite. We say that M accepts $\omega$ iff the computation is finite and the final configuration contains the state $q_{accept}$.

Informally, any *quantum computer* finds a certain result ($q = q_{accept}$) or not ($q = q_{reject}$) starting from a certain input $\omega$ for a finite number of configurations or not. Those three cases are disjunctively. In fact, the definition by Hilbert space T as above is a generalization as far as it admits to find a result ($q = q_{accept}$) not for a finite number of configurations. The generalization (unlike a properly arithmetical definition of Turing machine) needs "actual infinity" for the transfinite calculations to make sense:

New and new dimensions of T are added to the dimension of the finite input $\omega$ according to the program unambiguously until a result be obtained ($q = q_{accept}$) or not ($q = q_{reject}$). A calculation being defined by Hilbert space T can continue as if "in infinity" where it also can obtain a result ($q = q_{accept}$) or not ($q = q_{reject}$). If any Turing machine makes that calculation, it would need an infinite number of configurations and thus, infinite time. However, if the same calculation is made by a quantum computer, it can finish in a certain (even zero, theoretically) time. Once a certain result of that calculation whether ($q = q_{accept}$) or not ($q = q_{reject}$) is delivered by a quantum computer, it can be checked by a Turing machine for some polynomial time("non-P, but NP) or not ("non-P and non-NP"): the paper demonstrates that the former class is not empty.

*Quantum computer differs from Turing machine by actual infinity, which is allowed for the former. Anyway, the use of actual infinity is restricted only to the alphabet in the definition of quantum computer. The restriction relies on the intermediate stage of the redefinition of Turing machine by an infinitely dimensional vector space on a finite field. That finite field is transformed into an infinite field in the ultimate definition of quantum computer and allows for the aspects of arithmetic (finite) and set theory (actual infinite) to be both unified and disjunctively divided: the finite, arithmetical aspect is referred to the dimensionality of the vector space, and the actually infinite, set theory aspect to the field, on which the vector space is defined. An eventual appearance of actual infinity in the steps and configurations of quantum computer can be transferred always into the values of qubits, thus conserving the former finite, and the corresponding calculation as accomplishable for a certain finite time, theoretically even zero.*